# A 3D Printed Quad-Ridged Flared Horn Antenna Feeder for Radio-Telescopes

Andreas Hofmann[1], Yorgos Stratakos[2], Leonidas Marantis[2], Vasileios Spanakis-Misirlis[2], Danti Khouri[1], Athanasios Kanatas[2]

[1] Golden Devices GmbH, Erlangen, Germany, andreas.hofmann@golden-devices.com

[2] Department of Digital Systems, School of ICT, University of Piraeus, Piraeus, Greece, leomarantis@unipi.gr

*Abstract*—This study proposes a 3D-printed, quad-ridged, dual-pol, flared horn antenna feeder to meet the demands of a modern radio-telescope. The research work presented in this paper involves both the theoretical and the customized / adapted mechanical design of the proposed horn antenna, as well as a description of the 3D-printing and coating process. A novel slotted waveguide approach is followed for perfectly homogeneous coating. Additional mechanical adaptions (i.e. optimized feed section, radome) are employed to provide realistic simulation results. The proposed antenna offers a dual polarization capability and an attractive radiation beam to properly illuminate a 6m mesh dish reflector. A satisfying impedance matching from 1-3 GHz is achieved in both polarizations while maintaining an excellent isolation between the two N-type ports. Reflection coefficient and radiation pattern simulation results are presented, derived from the theoretical and the adapted mechanical design, exhibiting a close agreement.

## I. Introduction

Most of the fundamental astronomical advances achieved in the past sixty years have been made through radio astronomy, and many more are expected in the following years, since the radio astronomical observations provide the only way to investigate some cosmic phenomena. The ARGOS project is a concept for a leading-edge, low-cost European astronomical facility that will directly address several fundamental scientific questions, from the nature of dark matter to the origin of Fast Radio Bursts and the properties of extreme gravity, thereby satisfying urgent needs of the community.

In order to observe the deep space signals that ARGOS is looking for, a high-end front-end chain is developed. The first stage of that chain is the antenna feed (of the dish reflector), a custom-made antenna structure that plays a pivotal role in receiving the faint signals, efficiently collecting them, and funneling them into the receiver for further analysis. Through meticulous simulations and comparisons between different feed elements, a design has been made to satisfy the radio-telescope's demanding requirements. Factors like reflection coefficient response, radiation pattern and overall system efficiency are evaluated for the best performance of the feed.

Most of the conventional metal horn antennas operating around 1-3 GHz exhibit significantly large dimensions [1][2]. They are usually manufactured by aluminum through a metal CNC machinery approach that is cost-effective and leads to extremely heavy antenna structures. The scope of the paper centers around a holistic approach to additive manufacturing, leveraging the combined benefits of 3D printing and metalization in order to (a) create lightweight structures, (b) reduce production time and (c) achieve lower fabrication costs. By optimizing the feed design and fabrication, the full potential of the radio-telescope can be unlocked. This translates to increased sensitivity, allowing us to detect fainter signals. A wider field of view enables us to observe a larger swathe of the sky (ideal for Fast Radio Bursts detection).

Section II presents the proposed dual-pol, quad-ridged flared horn antenna design along with the simulation results. In Section III, the customized design adaptions are demonstrated along with detailed simulation results. Section IV describes the novel fabrication details (3D-printing with non-radiating slots and coating) of the antenna prototype. Section V provides the final conclusions of the investigation.

## II. Antenna Configuration

### A. Antenna Design

The antenna feed is a custom microwave antenna horn which supports dual polarization in the frequency range of 1-3 GHz. Since the proposed design operates as a feeder of a parabolic dish reflector, it is necessary to satisfy the reflector geometry and illuminate the dish correctly in order to avoid spill-over or under-illumination effects. In the ARGOS case, the parabolic reflector is a 6 m prime focus dish with an 0.38 F/D ratio. Two alternative horn designs were investigated at the preliminary design stage of the project (straight flare and bell-shaped flare). The bell-shaped horn was selected since it exhibits a more symmetric radiation beam with higher realized gain and lower sidelobes. Dual polarization is achieved by two orthogonal pairs of ridges (placed in the opposite sides of the flare) [2][3]. Two orthogonal monopole probes are designed at the starting point of the circular waveguide section in order to excite the two orthogonal polarized modes [4][5][6]. The detailed layout of the antenna is presented in Fig. 1, along with two cut-plane views that zoom in the monopole probes. The basic dimensions of the antenna design are given in Table I.

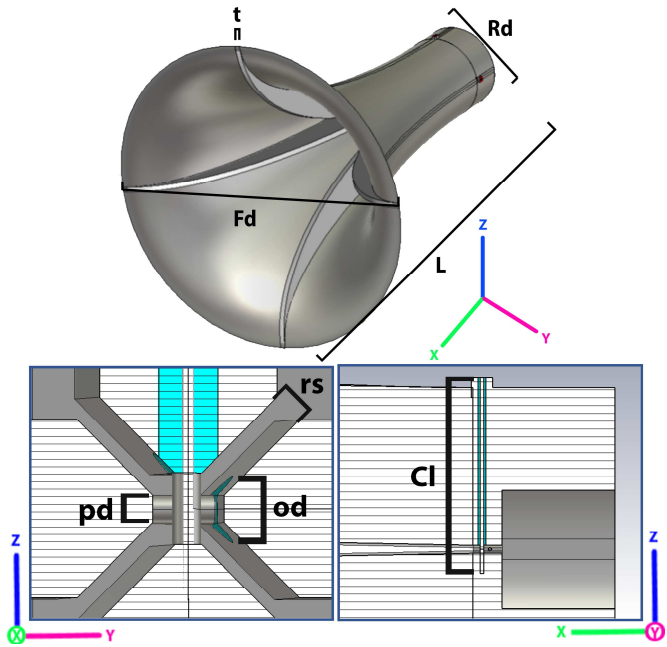

Fig. 1. The layout of the proposed quad-ridged bell-shaped horn simulation model in CST (top) and the two orthogonal monopole probes (bottom).

TABLE I. HORN ANTENNA DIMENSIONS

| | |
|---|---|
| Front diameter (Fd) | 445 mm |
| Rear diameter (Rd) | 124.03 mm |
| Total length (L) | 542.56 mm |
| Coaxial pin diameter (pd) | 0.914 mm |
| Coaxial outer diameter (od) | 2.11 mm |
| Coaxial line length (Cl) | 60.46 mm |
| Coaxial pins spacing (ps) | 1.31 mm |
| Ridges width (t - thickness) | 6.94 mm |
| Ridges spacing (rs) | 1.44 mm |

The feeder antenna has embedded N-type female coaxial ports which match accordingly the dual waveguide structure of the feed. While not the most compact option, N-type connectors demonstrate low losses and good performance.

*B. Simulation Results*

In this section the antenna feed is analyzed in the CST Microwave Studio simulator [7]. The feeder is operating in the frequency range from 1-3 GHz. The S-parameter plots, depicted in Figure 2, show that the probes themselves are not perfectly symmetrical (since they do not coincide). The isolation between the ports is excellent, exceeding -40 dB, signifying minimal signal leakage between them. Additionally, the reflection coefficients ($S_{11}$ and $S_{22}$) remain below -15 dB for most of the desired 1-3 GHz operational frequency range, indicating good impedance matching across the band. Figure 3 demonstrates the far-field radiation patterns for the realized gain in two orthogonal cuts (with 1 GHz step) for the bell-shaped flared horn antenna feed. The sidelobe levels remain below -15 dB for all patterns over the entire frequency range. The realized gain varies from 11.8 dB (at 1 GHz) to 14.2 dB (at 3 GHz). The beamwidth and the sidelobe level are optimized in order to eliminate both spill-over and under-illumination and achieve perfect illumination of the 6m parabolic dish reflector of the ARGOS radio-telescope.

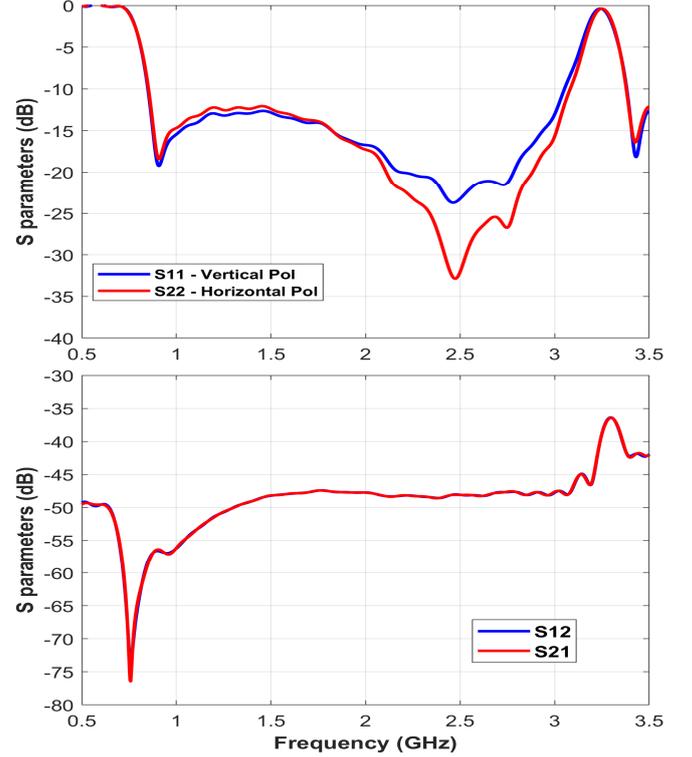

Fig. 2. S-Parameter simulation results of the theoretical horn model.

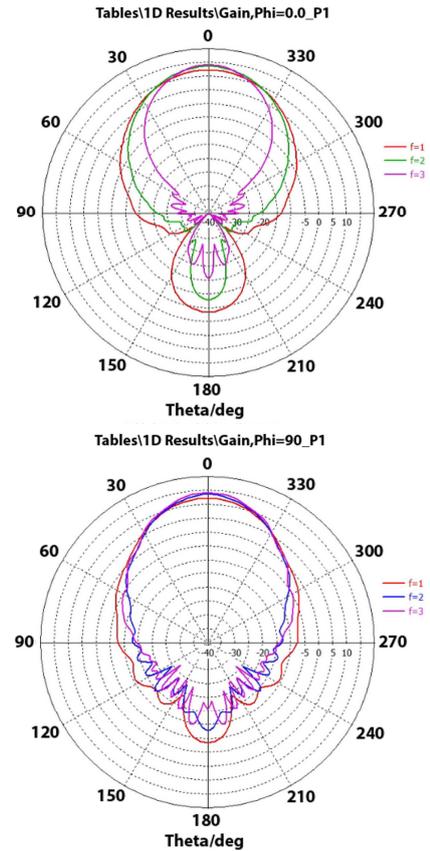

Fig. 3. Port 1 radiation patterns for 1, 2 and 3 GHz for φ = 0 and 90° cuts.

## III. DESIGN FOR MANUFACTURING

The antenna is manufactured by the metallization of 3D-printed polymer parts, which requires customized design adaptions. In order to account for the manufacturability and reproducibility of the manufacturing process, there are three main challenges. These are (a) the antenna feeding, (b) the homogeneous metal coating and (c) the radome design. The following subsections deal with the aforementioned issues.

### A. Antenna Feed

One of the basic challenges of the feeding-section of the antenna is the air-filled coaxial line for the monopole depicted in Fig. 1(bottom). The ratio of length to diameter of the inner conductor is very high. Moreover, it is not easy to have supports of the inner conductor. This makes it tough to account for a uniform concentric guidance. To overcome this, a semi-rigid cable with an outer diameter of 0.141 inch is used as feeding line. The cable is compatible for a mounting with an N-type connector and a stripped end of the cable can directly be used as a monopole. However, this requires that the electrical contact between the cable and the antenna is ensured. Therefore, grub screws are used to ensure this contact. To maintain this approach, all simulations in this section are performed with small gaps between the cable and the antenna body, with the exception of the two grub screws. A sectional view of the semi-rigid feed section is shown in Fig. 4. The conductors of the semi-rigid cable are shown in green and the simplified grub screws in red, respectively.

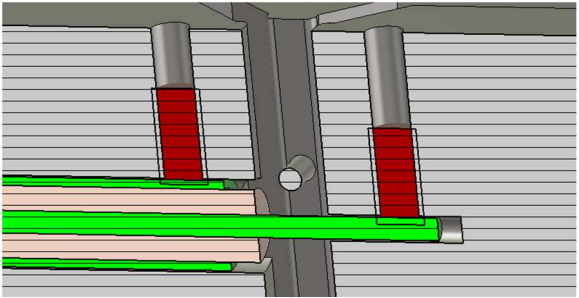

Fig. 4. Sectional view of the semi-ridig monopole probe in the feed section of the horn antenna.

The changes in the dimensions of the feed including the monopole alter the S-parameters. Moreover, it is not possible to print ridges with infinitely thin edges as shown in Fig. 1. Therefore, the geometry of the ridges was further optimized including rounded edges on the inner sides of the ridges to ensure a proper matching and a good coincidence between measurement and simulation results. The optimized ridge geometry including the dimensions is depicted in Fig. 5. The corresponding reflection loss simulation results (Fig. 6) are in good agreement with the results of the theoretical antenna model (Fig. 2). $S_{11}$ plots are well below -10 dB over the desired frequency range from 1-3 GHz. There is even a safety margin on the upper and lower frequency to account for manufacturing tolerances.

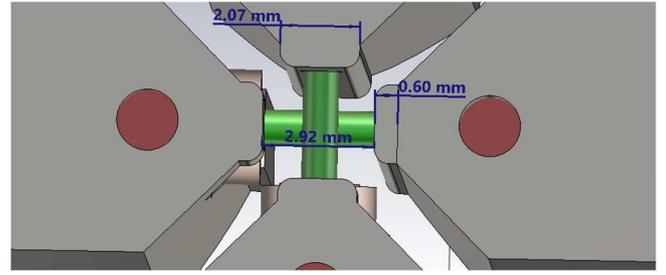

Fig. 5. Optimized ridge geometry demonstrating rounded edges.

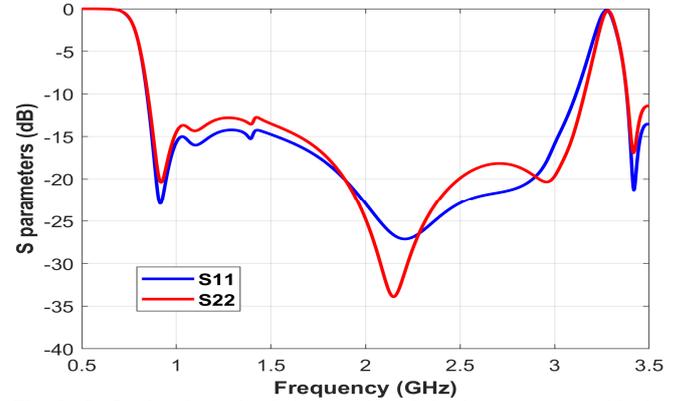

Fig. 6. Reflection loss simulation results of the horn antenna with the optimized feed section (semi-rigid cable & rounded edge ridges).

### B. Homogenous coating

In order to allow for a closed and homogenous coating of the inner walls of the antenna, while maintaining a quasi-monolithic design and manufacturing process, the novel slotted waveguide approach [8] is utilized. Usually, slots are introduced in waveguides to allow for the coating of the inner waveguide walls. These slots are non-radiating, since they are smaller than a quarter of the guided wavelength and are inserted in a manner, that no surface currents are cut perpendicular. In this specific case of the quad-ridged flared horn antenna feeder, the shortest wavelength is 10 cm. To ensure a homogenous coating slot sizes of 7 mm and a slot distance of 5 mm are sufficient and were chosen for the final design. With a dual polarized structure, it is not possible to use long slots as shown in [8]. Instead, novel rhomboid (in flare part) and rectangular openings (feeding part) are used as a substitute for slots. The slotted horn antenna is depicted in Fig. 7, and the corresponding reflection loss in Fig. 8.

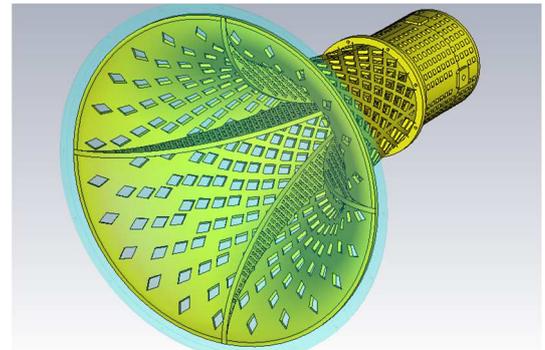

Fig. 7. Slotted bell-shaped horn with rhomboid and rectangular openings.

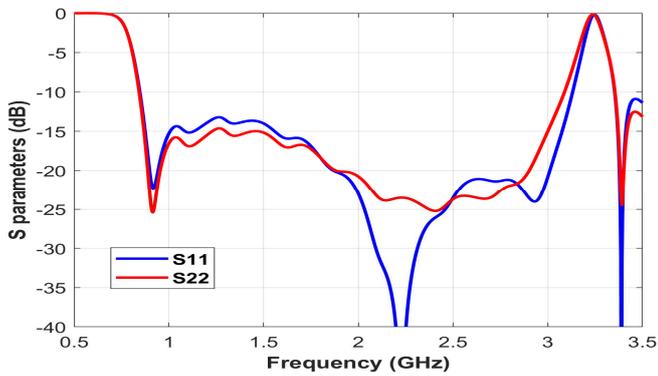

Fig. 8. Reflection loss of the slotted horn antenna.

As expected, there are no major differences between the reflection loss of the slotted antenna (Fig. 8) and the non-slotted antenna (Fig. 6).

*C. Radome*

The feed horn will be installed and used outdoors. It is therefore exposed to a wide range of environmental conditions. In order to prevent water or dirt from entering the antenna, it is necessary to seal the antenna properly. The radome performs this task for the radiating aperture. To prevent a major distortion of radiation, the radome is chosen to maintain a spherical shape, centered on the phase center of the antenna. The phase centers of the antenna are simulated for different frequencies. The phase-center over frequency is quite stable its average origin is around 20 cm behind the center of the radiating aperture. The thickness of the radome is chosen at 4 mm to ensure mechanical stability and at the same time keep the losses at a minimum. The material used is a weather- and uv-resistant thermoplastic with a permittivity of 2.4. Simulations show a deterioration of the S-parameters due to the radome. To counteract this deterioration, the radome is not printed completely filled, but made of two outer layers to ensure tightness and a partially filled infill. This allows for the adjustability of effective permittivity. Since the infill is very small compared to the wavelength, simulations with an effective permittivity are sufficient. An effective permittivity of 2 yields a good result between the S-Parameter deterioration and the density of the infill. A full view of the radome mounted on the feed antenna is shown in Fig. 9.

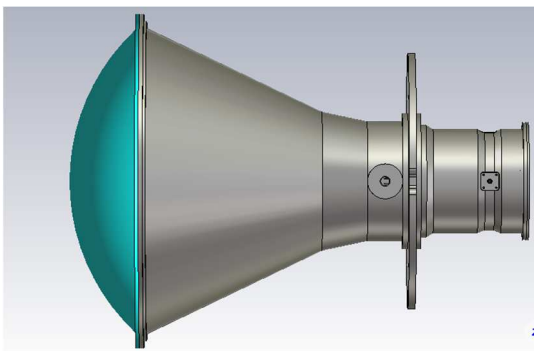

Fig. 9. Full view of the thermoplastic radome of the horn antenna.

The simulated reflection loss of the antenna with the mounted radome is depicted in Fig. 10. Due to the radome, the reflection loss exhibits a greater ripple effect compared to Fig. 8. However, the basic shape of the plot does not differ and the specification of the antenna is well fulfilled with an effective permittivity of 2.

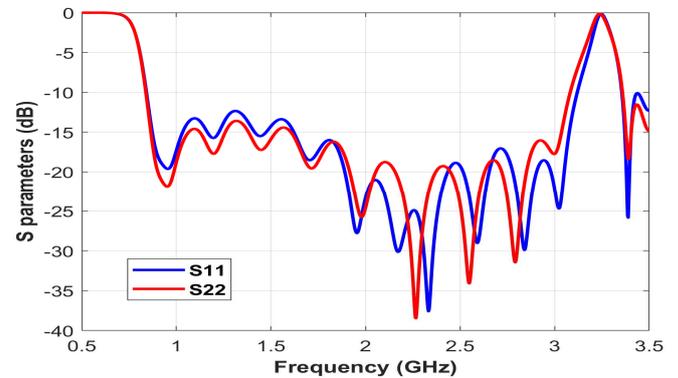

Fig. 10. Reflection loss of the horn antenna with the mounted radome.

The isolation of the two antenna ports is shown in Fig. 11. The isolation is well below -40 dB, it peaks slightly above -50 dB in the intended frequency range.

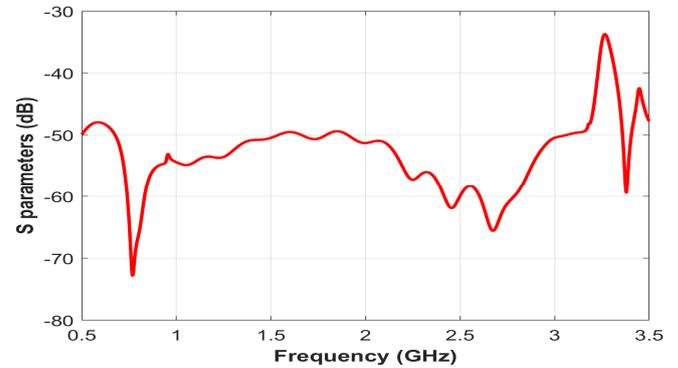

Fig. 11. Isolation of the feeding ports of the antenna with the radome.

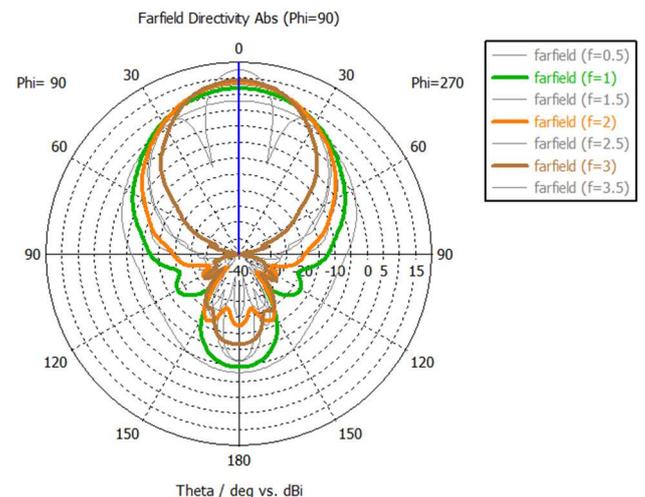

Fig. 12. Port 1 horizontal radiation pattern with the mounted radome.

Another critical factor of the feed-antenna is the beam-pattern. The simulated horizontal radiation patterns of port 1 are exemplary shown in Fig. 12. They also inhibit only minor deviations compared to Fig. 3. The angular 3dB width is only slightly increased by 2 degrees. The main lobe magnitude features a slight decrease of 0.3 dB. However, this is to be expected due to the additional losses caused by the radome.

## IV. Antenna Integration/ Fabrication Process

### A. 3D-Printing - Partitioning

The feed antenna must be mounted on a support ring. Therefore, it has to be partitioned in different parts. In addition, various parts of the antenna are predestined for production using different 3D printing processes. Moreover, the antenna is required to be weather-proof. Therefore, a weather-proof cover is also required to be mounted on the slotted parts of the antenna. The fabricated prototype of the antenna is shown in Fig. 13, along with the partitioning and the simplified parts.

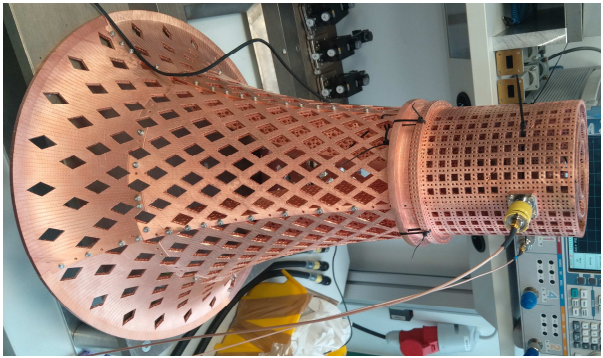

Fig. 13. Partitioning of the prototype horn antenna.

The short of the feeding section (cover at the back) must be removable for the installation of the semi-rigid cables. Since these parts have a rather simple geometry without sharp overhanging structures and tolerances of the parts, they are not particularly critical and are therefore best suited for production using fused deposition modeling (FDM). In addition, the use of FDM for the radome allows a variable filling density to adjust the required effective permittivity.

Things look different for the other two parts of the antenna. The feeding section (with the small rectangular slots) requires a high printing resolution to obtain small tolerances in order to meet the S-Parameter specification of the antenna. To account for this, a resin-based stereolithography printing (SLA) is employed. This method offers high resolution and low tolerances for the feeding part. Usually, high resolution printers feature a lower size of the building platform. However, by partitioning of the antenna SLA can be employed for the feeding section. In contrast, the flare part of the antenna is quite large, but the required tolerances are also released. This part is manufactured by Selective-Laser-Sintering (SLS).

As mentioned before, a closed cover will be mounted to ensure the functionality outdoors (see Fig. 9). The parts are screwed together and sealing rings are employed to ensure the tightness. A sectional view of an exemplary sealing joint between the radome cover and the antenna cover is depicted in Fig. 14. The sealing ring, shown in green, is clamped in mutual deepenings of the radome (in blue) and the antenna cover in grey. The opening for a screw with a threaded insert on the opposite side is also visible. To ensure optimal performance, a silver coating will be applied – opting for silver over gold for cost-effectiveness and copper's susceptibility to oxidation. The measured weight of the prototype feed antenna is only 5.5 kg (including the radome).

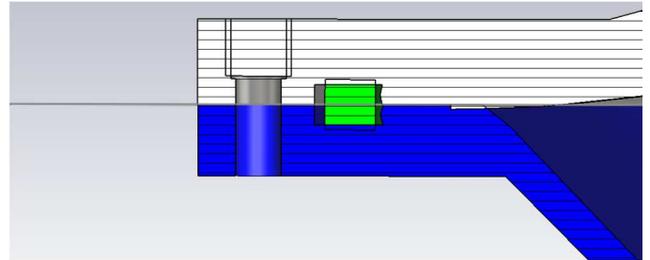

Fig. 14. Sectional view of a exemplary sealing joint

## V. Conclusion

In this work, a new dual-pol, quad-ridged flared horn antenna design was modelled and mechanically adapted in order to be fabricated by a novel 3D-printing approach. Novel non-radiating slots are adapted in order to provide a perfectly homogeneous coating. An optimized feed section and a special radome are employed to offer a realistic mechanical adaptation of the antenna model and better simulation results. Satisfying simulated $S_{11}$ plots and far-field pattern results were produced. Following this, the antenna will be fabricated and the feed will undergo strict experimental verification, including measurements of the $S_{11}$ versus frequency and characterization of the far-field radiation pattern.

### Acknowledgment

This research was funded partially by the University of Piraeus Research Center (UPRC) and the European Commission under the project ARGOS-CDS; Grant Agreement No. 101094354.